\documentclass[aps,prl,twocolumn,showpacs,preprintnumbers,superscriptaddress,groupedaddress]{revtex4}
\usepackage{graphicx}
\usepackage{dcolumn}
\usepackage{bm}
\usepackage{amsmath}
\usepackage{amssymb}
\usepackage{latexsym}
\usepackage{epsfig}
\usepackage{amsbsy}
\usepackage{array}
\usepackage{amssymb}
\usepackage{setspace}
\usepackage{bm}
\usepackage{url}

\begin{document}

\title{A criterion of the non-existence of surface states in a semi-infinite
crystal}
\date{\today }

\begin{abstract}
An infinite crystal can be constructed by an infinite number of parallel
two-dimensional (hkl) crystal planes coupled to each other. For crystals
with negligible spin-orbit coupling, we report a rigorous proof of a
criterion on the non-existence of surface states in a semi-infinite crystal
with the crystal symmetry. The forward transfer to be the same as the
backward one, called as F-B dynamical symmetry, is key to realize the
criterion. Based on lattice model Hamiltonian with coupling between the
nearest neighbor crystal planes only, we prove that a cut crystal will not
be able to accommodate any surface states if the original infinite crystal
has reflection symmetry about every crystal plane which results in F-B
symmetry. The criterion provide a platform to simply conclude whether
surface states exist or not in a cut crystal. For any such crystals, the
non-existence or existence of surface states depends on the cut direction of
the crystal plane. Since the spin-orbit coupling breaks the chiral symmetry,
resulting in the F-B asymmetry, surface states can emerge in the (hkl) cut
crystal with spin-orbit coupling.
\end{abstract}

\author{Huiping Wang}
\affiliation{State Key Laboratory of Surface Physics and Department of Physics, Fudan
University, Shanghai 200433, China}
\author{Tingting Gao}
\affiliation{State Key Laboratory of Surface Physics and Department of Physics, Fudan
University, Shanghai 200433, China}
\author{Ruibao Tao}
\altaffiliation[Corresponding author: ]{rbtao@fudan.edu.cn}
\affiliation{State Key Laboratory of Surface Physics and Department of Physics, Fudan
University, Shanghai 200433, China}
\maketitle

%\PACS{ }

\allowdisplaybreaks[4]

\textit{Introduction.---}Edge or surface, interface states, possessing some
novel physical properties, have been attracting considerable attention.
Recent decades have witnessed great interest in the study of edge or surface
states for 2D or 3D crystals. Many interesting and prominent physical
phenomena are tightly related to the existence of edge or surface states,
such as quantum Hall effect(QHE)\cite{Thouless-1982,Haldane-1988}, quantum
spin Hall effect(QSHE)\cite{Kane-Mele-2005, BernevigScience-2006, Liu-2008}
, topological insulator(TI) \cite%
{L.Fu-2007,Hsieh-2008,H.Zhang-2009,Y.L.Chen-2009,Hasan-2010,Qi-2011},
topological superconductor (TSC)\cite{Fu-Kane-2008,
Qi-2009,Das-2012,Deng-2012-2014} and topological Anderson insulator (TAI)
\cite{J.Li-2009, W.Li-2011, Y.Y.Zhang-2012} as well as surface
reconstruction of some semiconductors \cite{Northrup-1982, Fan-1989}. The
surface reconstruction in semiconductors Si and Ge can be ascribed to the
existence of surface states which provide the energy levels to be partly
filled with mobile surface charges coupled with surface softened phonon
modes. Gapless edge or surface states that exist in QSHE and TIs come from
the chiral symmetry breaking due to spin-orbit coupling (SOC), which is
highly attractive in recent studies.

In this letter,\emph{\ }we would like to focus on the majority of crystals
where the SOC is unimportant and can be neglected, such as some dielectric
materials like $ABO_{3}$ oxides, etc. For semiconductors or insulators,
surface states created by cutting the surface can provide some new physical
phenomenon. Especially, the electric conduction along domain walls in
ferroelectric materials has attracted intense recent studies\cite%
{Seidel-2009, Meier-2012} due to the possibility of creating and controlling
Nano-scale 1D/2D conductive paths in wide band gap insulators. In general,
for such insulating materials with negligible SOC, different cut surface of
the same crystal may show different behavior for the existence of surface
states. For experimentalists, it would be very much useful if there is a
criterion that can qualitatively tell which cut direction can be favorable
to generate much more surface states. The criterion may demonstrate the
underlying relationship between the existence or non-existence of surface
states and the crystal symmetry in the absence of SOC.

In general, an infinite 3D crystal can be described by an infinite number of
parallel two-dimensional crystal planes (CPs) which are periodically
arranged one by one with coupling. The direction of CPs can be denoted by
Miller indices (hkl), where h, k and l can be arbitrary integers. A
semi-infinite crystal with the (hkl) cut surface is called the (hkl) cut
crystal. In terms of a general lattice model Hamiltonian with the hopping
between the nearest neighbor (n.n.) CPs, the criterion can be phrased as
follows:

\textbf{Criterion: }A (hkl) cut crystal with negligible SOC will not allow
to have any surface states if the original infinite crystal has\ reflection
symmetry for every (hkl) crystal plane.

The criterion also covers the case of 2D crystals, then the "surface" just
means the edge line (atomic chain). Although the hopping are considered for
the n.n. CPs, the coupling within CPs can contain hopping to all possible
neighbors, i.e., not only the nearest neighbor (n.n.) ones. In our
discussion, the transfer matrix approach\cite{Lee1-1981,YYZhao-1} is applied.

The crystals with the reflection symmetry are only one of the two types:
Type I : \textquotedblleft $\cdots $-$P$-$P$-$P$-$P$-$\cdots $%
\textquotedblright\ [Fig.1a]\ and Type II: \textquotedblleft $\cdots $-$P$-$%
Q $-$P$-$Q$-$\cdots $\textquotedblright\ [Fig.1b]\ where $P\ $and $Q$
represent CPs. The same $P$ $(Q)$ represents exactly the same CP and $Q\neq
P $ means that $P$ and $Q$ are two different CPs. The bar \textquotedblleft
-\textquotedblright\ roughly describes the distance between the n.n. planes.
The same \textquotedblleft -\textquotedblright\ means the same distance.
Since Type II:\textquotedblleft $\cdots $-$P$-$Q$-$P$-$Q$-$\cdots $%
\textquotedblright \emph{\ }can be dynamically transformed into Type I:
\textquotedblleft $\cdots $-$P$-$P$-$P$-$P$-$\cdots $\textquotedblright ,
thus we concentrate our attention on the proof of the criterion for Type I
at first and then turn back to Type II.
\begin{figure}[th]
\includegraphics[width=8cm]{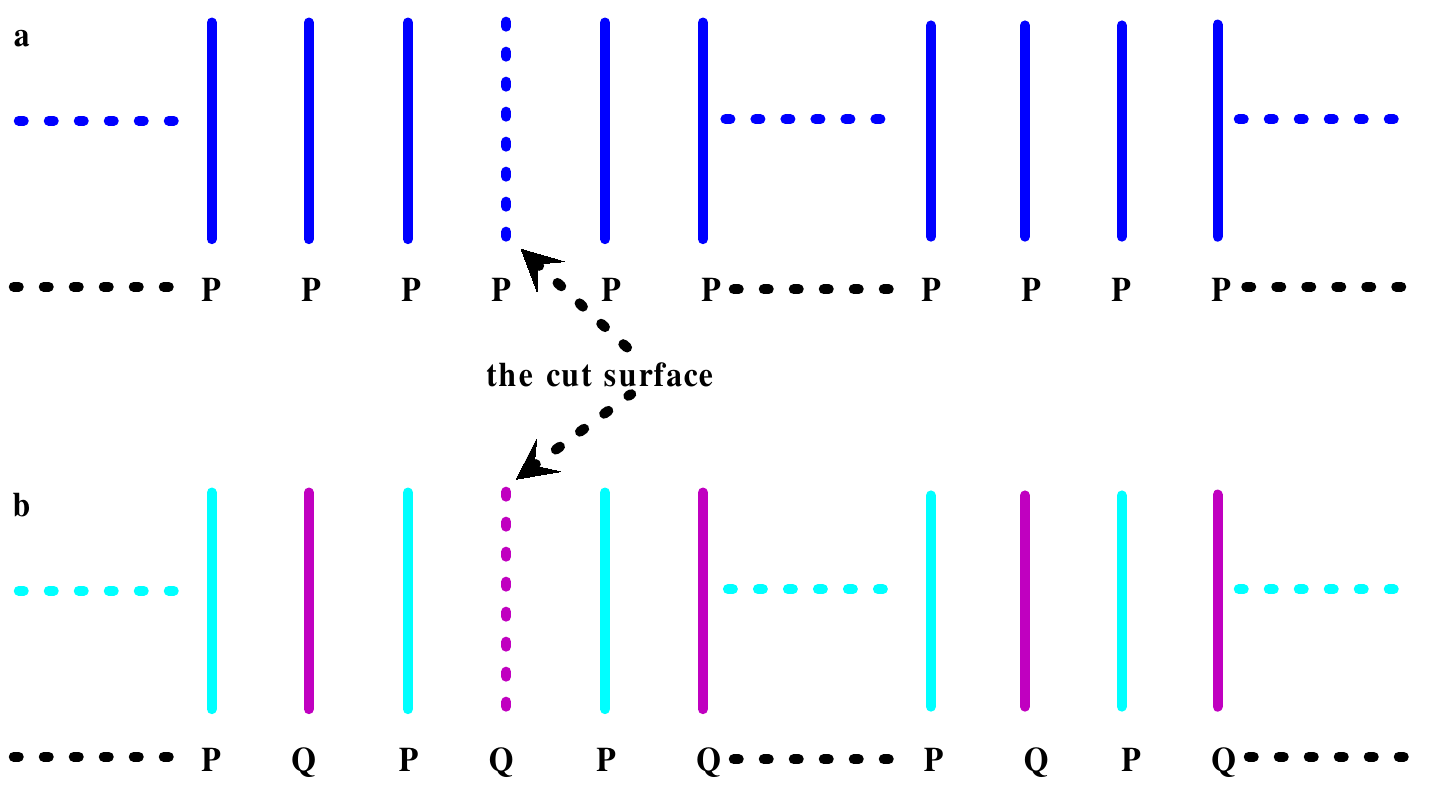}
\caption[Figure]{(color online). Two type structures. (a) Type I:
\textquotedblleft $\cdots $-$P$-$P$-$P$-$P$-$\cdots $\textquotedblright .
(b) Type II: \textquotedblleft $\cdots $-$P$-$Q$-$P$-$Q$-$\cdots $%
\textquotedblright .}
\label{twotypes}
\end{figure}

\textit{Proof of criterion for Type I.---}For the simplest case, each CP
only has single electron mode that corresponds to one atomic orbit per unit
cell. Under the n.n. hopping approximation between CPs, the study of surface
states in this case is exactly the same as that of edge states in the
semi-infinite 1D single orbit atomic chain. As is well known, no edge states
exist in the semi-infinite 1D atomic chain for both Type I and II when the
forward hopping constant equals to the backward one\cite{YYZhao-1}. Thus, we
will take into account the proof of the criterion for the case where each CP
$P$ contains $n$ $(>1)$ electron modes, that is, each CP contains many ( and
maybe different) atoms per unit cell and each atom may contribute many
different atomic orbits. For a semi-infinite crystal (SIC), each CP has the
periodic structure and is dimensional-wise lower than the original crystal.
Thus, the Fourier transformation (FT) is applied to each{\tiny \ }CP since
the wave vector $\overrightarrow{k}_{\parallel }$ is a good quantum number.
Take the diagonal representation of the Hamiltonian of each CP, then the
coupling between the n.n. CPs are introduced.

For a cut crystal \textquotedblleft $P$-$P$-$P$-$P$-$\cdots $%
\textquotedblright\ in Type I [Fig.1a], it is not difficult to obtain QDEs:
\begin{equation}
E_{n\times n}\Psi _{i}=F_{n\times n}\Psi _{i+1}+B_{n\times n}\Psi
_{i-1},i\geq 1;  \label{1}
\end{equation}%
here $\Psi _{i}^{T}=(\psi _{i}^{(1)}(\overrightarrow{k}_{\parallel }),\psi
_{i}^{(2)}(\overrightarrow{k}_{\parallel }),\cdots ,\psi _{i}^{(n)}(%
\overrightarrow{k}_{\parallel }))$ and \{$\psi _{i}^{(\alpha )}(%
\overrightarrow{k}_{\parallel }):\alpha \in \{1,2,\cdots ,n\}$\} are the
second quantized Fermi wave functions of $\alpha ^{th}$ electron mode in the
$i^{th}$ CP. $F_{n\times n}$ ($B_{n\times n}$) describes the $n\times n$
forward (backward) hopping matrix from the plane $P_{i}$ to its n.n. plane $%
P_{i+1}$ ($P_{i-1}$) and $n$ can be any finite positive integers. $%
B_{n\times n}=F_{n\times n}^{+}$ due to the Hermiticity of the Hamiltonian
and $\Psi _{0}=0$ is the boundary condition. $E_{n\times n}$ is defined as $%
E_{n\times n}=diag\{E_{1},E_{2},\cdots ,E_{n}\}$ and \{$E_{\alpha }=E-\omega
_{\alpha }(\overrightarrow{{k}}_{\parallel }):\alpha =1,2,\cdots ,n$\}. $E$\
are the energy levels of electron waves propagating in the SIC and $\{\omega
_{\alpha }:\alpha =1,2,\cdots ,n\}$ are energies of eigen-modes renormalized
at each CP. The elements of $F_{n\times n}$, $B_{n\times n}$ and $\{\omega
_{\alpha }:\alpha =1,2,\cdots ,n\}$ are $\overrightarrow{k}_{\parallel }$
dependent. From now on, we omit the symbol $\overrightarrow{k}_{\parallel }$
for simplicity. When the original infinite crystal has the reflection
symmetry for each crystal plane, then we have $F_{n\times n}=B_{n\times
n}=F_{n\times n}^{+}$. Eq.(1) can be rewritten as%
\begin{equation}
\begin{cases}
E_{n\times n}\Psi _{i}=F_{n\times n}\Delta \Psi _{i}, \\
\Delta \Psi _{i}=\Psi _{i+1}+\Psi _{i-1},%
\end{cases}
\label{2}
\end{equation}%
\ The matrices $F_{n\times n}$\ and $E_{n\times n}$\ are hermitian with
dimensionality $n$. Here we adopt the dimensional reduction method to reduce
the dimensionality $n$ in Eq. (2) to $1$. We will prove that no surface
waves can accommodate in a SIC for any energy $E$. For the general case, we
assume that $F_{n\times n}$ and $E_{n\times n}$ are arbitrary square
matrices and not limited to hermitian matrices.

\textit{Proof for }$n\geq 3$\textit{\ in Type I.}\textbf{---}By means of
dimensional reduction method, we will reduce the dimensionality $n$ in
Eq.(2) into $1$ or $2$. Let us first consider an energy such that $E:\det
(E_{n\times n})\neq 0$. Since $\det (E_{n\times n})\neq 0$, we can obtain
from Eq.(2)%
\begin{equation}
\Psi _{i}=\left( E_{n\times n}^{-1}F_{n\times n}\right) \Delta \Psi
_{i},i\geq 1.  \label{3}
\end{equation}%
It is well known that a square matrix $\left( E_{n\times n}^{-1}F_{n\times
n}\right) $ can be decomposed into a Jordan matrix via a similarity
transformation $J_{n\times n}=U_{n\times n}^{-1}\left( E_{n\times
n}^{-1}F_{n\times n}\right) U_{n\times n}$. $J_{n\times n}$ is a block
diagonal matrix: $J_{n\times n}=\sum_{i=1}^{s\oplus }J_{i}(\lambda _{i})$
where
\begin{equation*}
J_{i}(\lambda _{i})=\left(
\begin{array}{cccc}
\lambda _{i} & c_{i} & 0 & 0 \\
0 & \lambda _{i} & \ddots & 0 \\
0 & 0 & \ddots & c_{i} \\
0 & 0 & 0 & \lambda _{i}%
\end{array}%
\right) ,c_{i}\neq 0\text{ or }c_{i}=0,
\end{equation*}%
and $\{\lambda _{i}:i=1,2,\cdots ,s\}$ in $J_{n\times n}$ are eigenvalues of
the matrix $\left( E_{n\times n}^{-1}F_{n\times n}\right) $. Now we have
\begin{equation}
\Psi _{i}^{\prime }=J_{n\times n}\Delta \Psi _{i}^{\prime },\Psi
_{i}^{\prime }=U_{n\times n}^{-1}\Psi _{i},\Psi _{0}^{\prime }=0.  \label{4}
\end{equation}%
In terms of the lowest-right-most element of the Jordan matrix $J_{n\times
n} $, the first equation can be reached immediately for $\psi
_{i}^{(n)\prime }$
\begin{equation}
\psi _{i}^{(n)\prime }=\lambda _{s}\Delta \psi _{i}^{(n)\prime }.  \label{5}
\end{equation}%
Eq.(5) is exactly the same as the transfer matrix equation of 1D atom chain
with single electron mode. It has been known that there is no edge states
for any energy $E$ no matter whether $\lambda _{s}$ $=0$ or $\lambda
_{s}\neq 0$\cite{YYZhao-1}. Thus we arrive at \{$\psi _{i}^{(n)\prime
}=0:i\geq 1$\} for the solution of surface states. After back-substituting \{%
$\psi _{i}^{(n)\prime }=0:i\geq 1$\} into Eq.(4), we find that surface
states are also impermissible for the $(n-1)^{th}$ mode, yielding \{$\psi
_{i}^{(n-1)\prime }=0:i\geq 1$\}. After step by step, we obtain trivial
solutions of all surface waves: \{$\Psi _{i}^{\prime }=0:i\geq 1$\} that
results in $\Psi _{i}=U_{n\times n}\Psi _{i}^{\prime }=0$. Hence no surface
states are allowed for $\det (E_{n\times n})\neq 0$.

Next let us think over some energies $E$ such that: $\det (E_{n\times n})=0$%
. Now we apply a Jordan transformation $V_{n\times n}$ to the square matrix $%
E_{n\times n}:$ $E_{n\times n}^{J}=V_{n\times n}^{-1}E_{n\times n}V_{n\times
n}$ and have%
\begin{equation}
E_{n\times n}^{J}\Psi _{i}^{\prime }=F_{n\times n}^{\prime }\Delta \Psi
_{i}^{\prime },\Psi _{i}^{\prime }=V_{n\times n}^{-1}\Psi _{i},\Psi
_{0}^{\prime }=0,i\geq 1,  \label{6}
\end{equation}%
here $F_{n\times n}^{\prime }=V_{n\times n}^{-1}F_{n\times n}V_{n\times n}$.
$E_{n\times n}^{J}=\sum_{i=1}^{s\oplus }J_{i}(\lambda _{i})$ where we have
arranged such that the sub-matrix $J_{1}$ contains $\lambda _{1}=0$. Without
loss of generality, we can assume the first block is a two-order Jordan
sub-matrix at first. For other cases where $J_{1}(\lambda _{1}=0)$ is one or
greater than two, we can do similar demonstration as we do for a two-order
Jordan block $J_{1}(\lambda _{1}=0).$ The derivation can proceed by
considering two scenarios: 1) Suppose $F_{11}^{\prime }\neq 0$. We can
obtain from the first row of Eq.(6),%
\begin{equation}
\Delta \psi _{i}^{(1)\prime }=-\frac{1}{F_{11}^{\prime }}\sum_{\alpha
=2}^{n}F_{1\alpha }^{^{\prime }}\Delta \psi _{i}^{(\alpha )\prime
}+c_{1}\psi _{i}^{(2)\prime }:i\geq 1.  \label{7}
\end{equation}%
Substituting Eq.(7) into Eq.(6), we can arrive at%
\begin{equation}
E_{(n-1)\times (n-1)}^{(1)}\Psi _{i}^{\prime \prime }=F_{(n-1)\times
(n-1)}^{(1)}\Delta \Psi _{i}^{\prime \prime },  \label{8}
\end{equation}%
here $\Psi _{i}^{\prime \prime T}=(\psi _{i}^{(2)\prime },\psi
_{i}^{(3)\prime },\cdots ,\psi _{i}^{(n)\prime })$ and \{$F_{\alpha \beta
}^{(1)}=F_{\alpha \beta }^{\prime }-F_{\alpha 1}^{\prime }F_{1\beta
}^{\prime }/F_{11}^{\prime }$; $\alpha ,\beta \in \left\{ 2,3,\cdots
,n\right\} $ \}. Thus, we have reduced the dimensionality $n$ in Eq.(6) into
$n-1$.

2) Next suppose $F_{11}^{\prime }=0$: Now we focus on the $1^{st}$ column
matrix elements of $F_{n\times n}^{\prime }$. If all of \{$F_{j1}^{\prime
}:j=1,2,\cdots $ $,n$\} are zero, the reduction of the dimensionality in
Eq.(6) is already reached. Thus, we assume \{$F_{\beta 1}^{^{\prime }}$ $%
\neq 0:\beta \in \{2,3,\cdots ,n\}$\} without loss of generality and obtain%
\begin{equation*}
\Delta \psi _{i}^{(1)\prime }=-\frac{1}{F_{\beta 1}^{\prime }}\left(
\sum_{j=2}^{n}F_{\beta j}^{\prime }\Delta \psi _{i}^{(j)\prime }-\lambda
_{\gamma }\psi _{i}^{(\beta )\prime }-c_{\gamma }\psi _{i}^{(\beta +1)\prime
}\right) .
\end{equation*}%
After plugging the above equation into Eq.(6), we can get
\begin{equation}
\Theta _{(n-1)\times (n-1)}\Psi _{i}^{\prime \prime }=F_{(n-1)\times
(n-1)}^{\prime \prime }\Delta \Psi _{i}^{\prime \prime },i\geqslant 1,
\label{9}
\end{equation}%
here $\gamma $ $\in \{1,2,\cdots ,s\}$, $\{F_{lj}^{\prime \prime
}=F_{lj}^{\prime }-F_{l1}^{\prime }F_{\beta j}^{\prime }/F_{\beta 1}^{\prime
},j\in \{2,\cdots ,n\},l\in \{1,2,\cdots ,\beta -1,\beta +1,\cdots ,n\}\}$
and $\Psi _{i}^{^{\prime \prime T}}=(\psi _{i}^{(2)\prime },\psi
_{i}^{(3)\prime },\cdots ,\psi _{i}^{(n)\prime })$. Elements in the matrix $%
\Theta _{(n-1)\times (n-1)}$ are functions of energies \{$E_{\alpha }:$ $%
\alpha =1,2,\cdots ,n\}$ and hopping constants. $F_{(n-1)\times
(n-1)}^{\prime \prime }$ is a renormalized hopping matrix, dependent on the
energy $E$. As a result, the dimensionality $n$ in Eq.(6) has been reduced
to $n-1$. If the determinant of $E_{(n-1)\times (n-1)}^{(1)}$ or/and $\Theta
_{(n-1)\times (n-1)}$ is zero, we will continue to reduce the dimensionality
$n-1$ in Eq.(8) or/and Eq.(9) to $n-2$ by following the similar steps from
Eq.(6) to Eq.(9). If necessary, we can do more reductions similar to above
and eventually reduce the dimensionality in Eq.(6) to $1$ or $2$. Meanwhile,
it is easy to see that other modes $\{\Delta \psi _{i}^{(l)}:l=2,3,\cdots ,n$
or $l=3,4,\cdots ,n)\}$ are either the linear combinations of $\{\psi
_{i}^{(j)},\Delta \psi _{i}^{(j)}:j=1$ $($or $j=1,2)\}$ or can be decoupled
as local modes when the dimensionality in Eq.(6) is reduced to $1(2)$. No
surface states exist for the dimensionality $1$ (as well known) when the
forward hopping constant is equal to the backward hopping one, neither for
the dimensionality $2$, as will be proved in the following.

\textit{Proof for }$n=2$\textit{\ in Type I.}---When $n=2$, Eq.(2) is
rewritten as%
\begin{equation}
E_{2\times 2}\Psi _{i}=F_{2\times 2}\Delta \Psi _{i},\Psi _{0}=0,i\geq 1,
\label{10}
\end{equation}%
here $\Psi _{i}^{T}=(\psi _{i}^{(1)},\psi _{i}^{(2)})$ and assuming $%
E_{2\times 2}$ and $F_{2\times 2}$ are general matrices in order to cover
the previous case where the dimensionality in Eq.(6) is reduced to $2$ when $%
n\geq 3$ and $\det (E_{n\times n})=0$. To ensure the proof valid for any
energy $E$ and any crystal structures, we must discuss all possible matrix
structures of $E_{2\times 2}$ and $F_{2\times 2}.$

At first, note that when $\det (E_{2\times 2})\neq 0$ or $\det (F_{2\times
2})\neq 0$, we obtain \{$\psi _{i}^{(1)}=0,\psi _{i}^{(2)}=0:i\geq 1$\} for
surface waves by following the similar steps from Eq.(3) to Eq.(5).

Next, think over the special case where $\det (E_{2\times 2})=0$ and $\det
(F_{2\times 2})=0$. We apply a Jordan similar transformation $U_{2\times 2}$
for $E_{2\times 2}$, then Eq.(10) can be written as%
\begin{equation}
\begin{cases}
J_{2\times 2}^{\prime }\Psi _{i}^{\prime }=F_{2\times 2}^{\prime }\Delta
\Psi _{i}^{\prime },\Psi _{i}^{\prime }=U_{2\times 2}^{-1}\Psi _{i}, \\
\\
J_{2\times 2}^{\prime }=U_{2\times 2}^{-1}E_{2\times 2}U_{2\times 2}=\left(
\begin{array}{cc}
\lambda _{1} & c_{0} \\
0 & 0%
\end{array}%
\right) ,%
\end{cases}
\label{11}
\end{equation}%
here $F_{2\times 2}^{\prime }=U_{2\times 2}^{-1}F_{2\times 2}U_{2\times 2}$
and $\{\lambda _{1},c_{0}\}$ can be zero or nonzero. We further examine the
following three possible situations:

i) $c_{0}=0$ and $\lambda _{1}=0$

We can apply the Jordan transformation again to $F_{2\times 2}^{^{\prime }}$
and since $\det (F_{2\times 2})=\det (F_{2\times 2}^{\prime })=0$, Eq.(11)
becomes%
\begin{equation}
\begin{cases}
0_{2\times 2}=J_{2\times 2}^{\prime \prime }\Delta \Psi _{i}^{\prime \prime
},\Delta \Psi _{i}^{\prime \prime }=W_{2\times 2}^{-1}\Delta \Psi
_{i}^{\prime \prime }, \\
\\
J_{2\times 2}^{\prime \prime }=W_{2\times 2}^{-1}F_{2\times 2}^{^{\prime
}}W_{2\times 2}=\left(
\begin{array}{cc}
\alpha _{1} & c_{1} \\
0 & 0%
\end{array}%
\right) .%
\end{cases}
\label{12}
\end{equation}%
When $\alpha _{1}=0$ and $c_{1}=0$, \{$\Delta \psi _{i}^{(1)\prime \prime
},\Delta \psi _{i}^{(2)\prime \prime }$\} fully decouple and become local
modes within each CP. When $\alpha _{1}\neq 0$ and $c_{1}=0$, $\Delta \psi
_{i}^{(1)\prime \prime }=0$ corresponds to an extended mode and $\Delta \psi
_{i}^{(2)\prime \prime }$ is decoupled as the local mode. When $\alpha _{1}=0
$ and $c_{1}\neq 0$, $\Delta \psi _{i}^{(2)\prime \prime }=0$ means the
non-existence of surface states and $\Delta \psi _{i}^{(1)\prime \prime }$
becomes the local modes without propagation among the CPs.

ii) $c_{0}\neq 0$ and $\lambda _{1}=0$\ \ \ \ \ \

At first, we note that when $F_{21}^{^{\prime }}\neq 0$ or $F_{22}^{\prime
}\neq 0$, $\psi ^{(1)\prime }$ and $\psi ^{(2)\prime }$ become local modes
within each CP or are zero solutions for surface states.

Next, consider the special case where $F_{21}^{\prime }=0$ and $%
F_{22}^{\prime }=0$, then we get from Eq.(11)
\begin{equation}
c_{0}\psi _{i}^{(2)\prime }=F_{11}^{^{\prime }}\Delta \psi _{i}^{(1)\prime
}+F_{12}^{\prime }\Delta \psi _{i}^{(2)\prime }.  \label{13}
\end{equation}%
When $F_{11}^{\prime }=0$, Eq.(13) turns into $c_{0}\psi _{i}^{(2)\prime
}=F_{12}^{\prime }\Delta \psi _{i}^{(2)\prime }$ and we obtain \{$\psi
_{i}^{(2)\prime }=0:i\geq 1$\} for surface states. When $F_{11}^{\prime
}\neq 0$, $\{\psi _{i}^{(1)\prime },\psi _{i}^{(2)\prime }\}$ are coupled
together. If there are surface states existing for $\{\psi _{i}^{(1)\prime
},\psi _{i}^{(2)\prime }\}$, we can have $\Delta \psi _{i}^{(1)\prime
}=\beta \Delta \psi _{i}^{(2)\prime }$ where $\beta $ is a non-zero
constant. Then Eq. (13) becomes $c_{0}\psi _{i}^{(2)\prime }=(F_{11}^{\prime
}\beta +F_{12}^{\prime })\Delta \psi _{i}^{(2)\prime }$ that results in \{$%
\psi _{i}^{(2)\prime }=0:i\geq 1$\} for surface modes, leading to \{$\psi
_{i}^{(1)\prime }=0:i\geq 1$\}. Therefore, no surface states can exist in
the SIC.

iii) $c_{0}=0$ and $\lambda _{1}\neq 0$

The proof is almost exactly similar to the case \{$c_{0}\neq 0$ and $\lambda
_{1}=0$\} and we get the same conclusion.

Up to now, the criterion has been analytically proved for the cut crystals
with \textquotedblleft $Q$-$Q$-$Q$-$Q$-$\cdots $\textquotedblright\ by means
of dimensional reduction method.

\bigskip \textit{Proof of criterion for Type II.---}In Type II, the crystal
has two different CPs: $P$ and $Q$. We just discuss the $Q$ cut crystal
\textquotedblleft $Q$-$P$-$Q$-$P$-$\cdots $\textquotedblright\ [Fig.1b]
since the discussion for $P$ cut crystal will be similar. Now the QDEs for
the $Q$ cut crystal are%
\begin{equation}
\begin{cases}
E_{n\times n}^{(P)}\Psi _{i}=F_{n\times m}\left( \Phi _{i}+\Phi
_{i+1}\right) , \\
E_{m\times m}^{(Q)}\Phi _{i}=\left( F_{n\times m}\right) ^{+}\left( \Psi
_{i-1}+\Psi _{i}\right) ,i\geq 1,%
\end{cases}
\label{14}
\end{equation}%
here the boundary conditions are $\Phi _{0}=0_{m\times 1}$ and $\Psi
_{0}=0_{n\times 1}$ and the CP $P$ has $n$ modes and $Q$ has $m$ ones. $n$
and $m$ can be equal or unequal. \{$E_{n\times n}^{(P)},E_{m\times m}^{(Q)}$%
\} are defined as $E_{l_{\alpha }\times l_{\alpha }}^{(\alpha
)}=diag\{E_{1}^{\alpha },E_{2}^{\alpha },\cdots ,E_{l_{\alpha }}^{\alpha }\}$%
, \{$E_{i}^{\alpha }=E-\omega _{i}^{\alpha }$: $i=1,2,...l_{\alpha }$\} and $%
l_{P\left( Q\right) }=n\left( m\right) $ when $\alpha =P\left( Q\right) $.
After some simple calculations, Eq.(14) can be rewritten as
\begin{equation*}
\begin{cases}
E_{(n+m)\times (n+m)}\Pi _{i}=F_{(n+m)\times (n+m)}\Delta \Pi _{i}, \\
E_{(n+m)\times (n+m)}=\left(
\begin{array}{cc}
2E_{n\times n}^{(P)} & -2F_{n\times m} \\
-2\left( F_{n\times m}\right) ^{+} & E_{m\times m}^{(Q)}%
\end{array}%
\right) , \\
F_{(n+m)\times (n+m)}=\left(
\begin{array}{cc}
0_{n\times n} & 0_{m\times n} \\
\left( F_{n\times m}\right) ^{+} & 0_{m\times m}%
\end{array}%
\right) ,\Pi _{i}=\left(
\begin{array}{c}
\Psi _{i} \\
\widetilde{\Phi }_{i}%
\end{array}%
\right) ,%
\end{cases}%
\end{equation*}%
here $\widetilde{\Phi }_{i}=\Phi _{i+1}+\Phi _{i}$ and $\Delta \Pi _{i}=\Pi
_{i-1}+\Pi _{i+1}$. Now the $Q$ cut crystal \textquotedblleft $Q$-$P$-$Q$-$P$%
-$\cdots $\textquotedblright\ in Type II\ is equivalent to the structure
\textquotedblleft $\widetilde{Q}$-$\widetilde{Q}$-$\widetilde{Q}$-$%
\widetilde{Q}$-$\cdots $\textquotedblright\ in Type I with the
dimensionality $n+m$. We can readily find $\{\Pi _{i}=0$: $\widetilde{\Phi }%
_{i}=0,\Psi _{i}=0,i\geq 1\}$ for surface waves. \{$\widetilde{\Phi }%
_{i}=0:i\geq 1$\} yields \{$\Phi _{i+1}+\Phi _{i}=0:i\geq 1$\} that leads to
\{$\Phi _{i}=0:i\geq 1$\}. Hence the criterion is also valid for Type II. So
far, we have completed the proof of the criterion of the non-existence of
surface states in the cut crystals for Type I \textquotedblleft $\cdots $-$P$%
-$P$-$P$-$P$-$\cdots $ \textquotedblright\ and Type II \textquotedblleft $%
\cdots $-$P$-$Q$-$P$-$Q$-$\cdots $\textquotedblright .

From the demonstration above, we clearly know that $F_{n\times n}=$ $%
B_{n\times n}$ is the key point for the non-existence of surface states in
the cut crystal. Other crystal structures, like $``\cdots $-$P$=$Q$-$P$=$Q$-$%
\cdots \textquotedblright $, $``\cdots $-$P$=$P$-$P$=$P$-$\cdots
\textquotedblright $,\thinspace $``\cdots $-$P$-$Q$-$S$-$P$-$Q$-$S$-$\cdots
\textquotedblright $, etc, do break the reflection symmetry (F-B symmetry)
in the above criterion, thus surface states can emerge in the SIC and can
contribute some surface bands in the bulk band gap.

In application of the criterion, we can easily check the armchair edged
graphene does not have edge states, since it has "$P$-$P$-$P$-$\cdots $"
structure. The conclusion is consistent with previous theoretical analysis%
\cite{YYZhao-2}. However, the type structure of zigzag edged graphene is
\textquotedblleft $P$=$P$-$P$=$P$-$\cdots $\textquotedblright , where the
F-B symmetry is broken, thus it is in favor of the existence of edge states
according to our criterion. From structure symmetry analysis of Perovskite
structure $ABO_{3}$ materials such as $PbTiO_{3}$ in different phases, we
can easily conclude that the c-cut $ABO_{3}$ materials in the para-electric
phase and with the polarization normal to c-axis have no surface states
since their structures are the "$P$-$Q$-$P$-$Q$-$\cdots $" type and but the
c-cut $ABO_{3}$ materials with the polarization along the c-axis favor
surface states due to the reflection asymmetry[Fig.2a]. Furthermore for
hexagonal structure c-cut ferroelectric $YMnO_{3}$[Fig.2b], it have surface
states due to the F-B asymmetry, consistent with the previous analysis\cite%
{Meier-2012}. The conclusions of previous theoretical works\cite{Ho-1975,
Mostoller-1982} can be readily qualitatively understood from the criterion.
\begin{figure}[th]
\includegraphics[width=8cm]{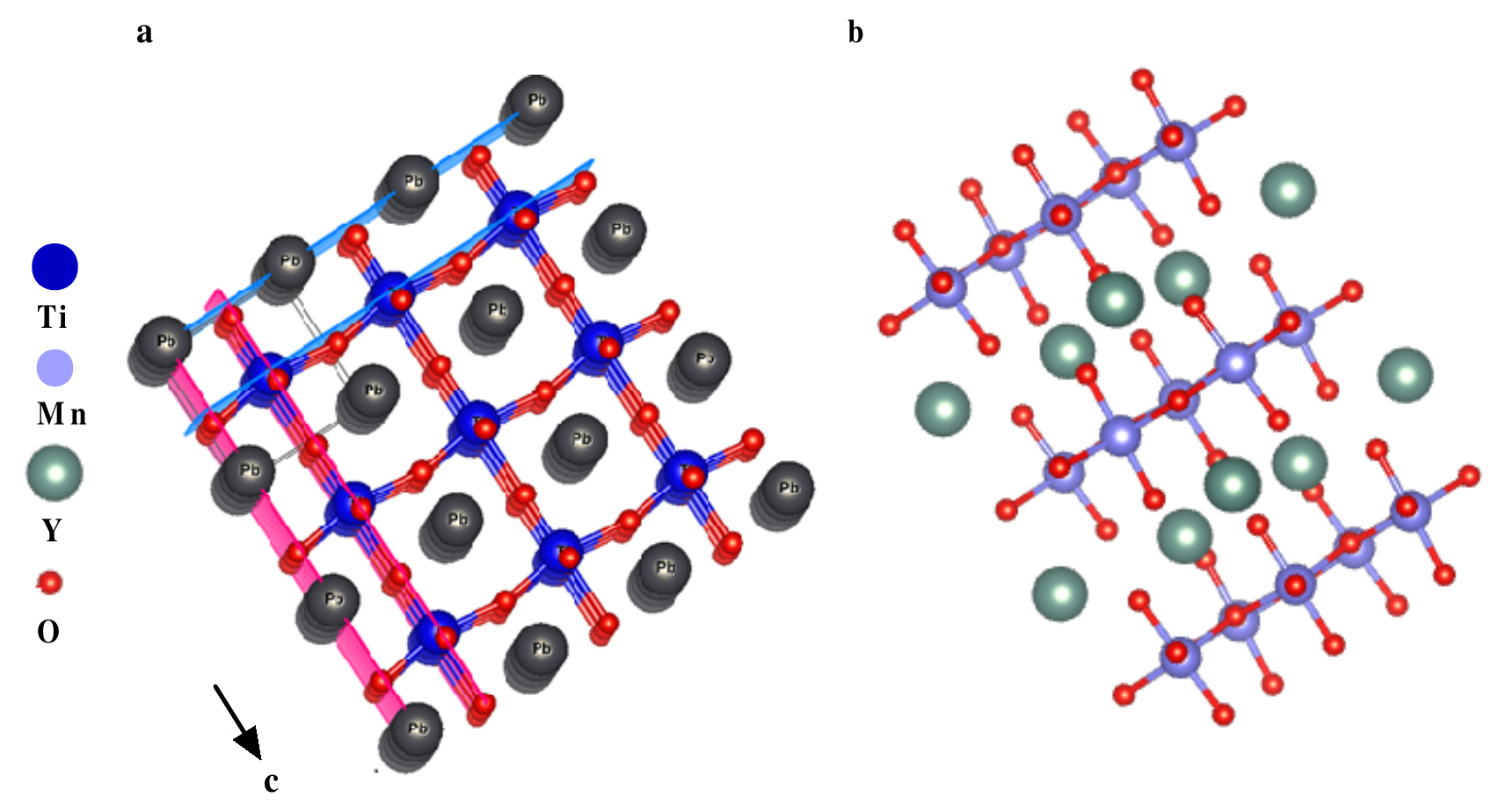}
\caption[Figure]{(color online). Crystal structures of c-cut $PbTiO_{3}$ and
c-cut $YMnO_{3}$ at the ferroelectric phases. (a) c-cut $PbTiO_{3}$ at the
ferroelectric phase. (b) c-cut $YMnO_{3}$ at the ferroelectric phase.}
\label{twostructures}
\end{figure}

\textit{Conclusion.---} We have rigorously proved a criterion on the
non-existence of surface states in a (hkl) cut crystal: there will not be
any surface states if the original infinite crystal has reflection symmetry
about every (hkl) crystal plane. In our demonstration, The longer range
hopping among CPs has not been considered and the many body correlation is
been neglected. Note the reflection symmetry is just a sufficient condition
for the non-existence of surface states in a cut crystal. In fact, the F-B
dynamical symmetry ($F_{n\times n}=B_{n\times n})$ is key to realize the
criterion. The F-B symmetry is more general and can be also applicable to
other structure crystals. For crystals with negligible SOC, one can find
that different cut surface of the same crystals may have different behavior
for the existence of surface states. While for crystals with SOC, such as
topological insulators, they break the chiral symmetry, resulting in the F-B
asymmetry, thus surface states can emerge.{\tiny \ } Moreover, the criterion
can be extended to $F_{n\times n}=e^{i\delta }B_{n\times n}$ where $\delta $
is a $\overrightarrow{k}$-dependent or zero. Much more detailed
investigations are underway for the longer rang hopping among CPs. We hope
the theoretical predication from our criterion will be helpful to determine
which cut direction of the crystals is in favor of generating surface modes
in new materials.

We gratefully acknowledge Professors Zhi-Xun Shen, Keji Lai, Tao Xiang, Xi
Dai, Zhong Fang and Yongliang Yang for{\tiny \ }helpful discussions. The
work is supported by the National Basic Research Program of China (973
Program) under the grant No.2011CB921803 and the National Natural Science
Foundation of China through the grant No.11147001.


\begin{thebibliography}{99}
\bibitem{Thouless-1982} D. J. Thouless, M. Kohmoto, M. P. Nightingale, and
M. den Nijs, Phys. Rev. Lett. 49, \textbf{405} (1982).

\bibitem{Haldane-1988} F. D. M. Haldane, Phys. Rev. Lett. \textbf{61}, 2015
(1988).

\bibitem{Kane-Mele-2005} C. L. Kane and E. J. Mele, Phys. Rev. Lett. \textbf{%
95}, 146802 (2005); C. L. Kane and E. J. Mele, Phys. Rev. Lett.\textbf{\ 95}%
, 226801 (2005).

\bibitem{BernevigScience-2006} B. A. Bernevig, T. L. Hughes, and S. C.
Zhang, Science \textbf{314}, 1757 (2006).

\bibitem{Liu-2008} C. X. Liu, T. L. Hughes, X. L. Qi, K. Wang, and S. C.
Zhang, Phys. Rev. Lett. \textbf{100}, 236601 (2008).

\bibitem{L.Fu-2007} L. Fu, C. L. Kane, and E. J. Mele, Phys. Rev. Lett.
\textbf{98}, 106803 (2007); L. Fu and C. L. Kane, Phys. Rev. Lett. \textbf{%
100}, 096407 (2008).

\bibitem{Hsieh-2008} D. Hsieh et al. Nature \textbf{452}, 970 (2008).

\bibitem{H.Zhang-2009} H. Zhang, C. X. Liu, X. L. Qi, X. Dai, Z. Fang, and
S. C. Zhang, Nat. Phys. \textbf{5}, 438 (2009).

\bibitem{Y.L.Chen-2009} Y. L. Chen et al., Science \textbf{325}, 178 (2009).

\bibitem{Hasan-2010} M. Z. Hasan and C. L. Kane, Rev. Mod. Phys. \textbf{82}%
, 3045 (2010).

\bibitem{Qi-2011} X. L. Qi and S. C. Zhang, Rev. Mod. Phys. \textbf{83},
1057 (2011).

\bibitem{Fu-Kane-2008} L. Fu and C. L. Kane, Phys. Rev. Lett. \textbf{100},
096407 (2008).

\bibitem{Qi-2009} X. -L. Qi, T. L. Hughes, S. Raghu, and S. -C. Zhang, Phys.
Rev. Lett. \textbf{102}, 187001 (2009); X.-L. Qi, T. L. Hughes, and S. C.
Zhang, Phys. Rev. B \textbf{82}, 184516 (2010).

\bibitem{Das-2012} A. Das, Y. Ronen, Y. Most, Y. Oreg, M. Heiblum, and H.
Shtrikman, Nat. Phys. \textbf{8}, 887 (2012).

\bibitem{Deng-2012-2014} S. Deng, L. Viola, and G. Ortiz, Phys. Rev. Lett.
108, 036803 (2012); S. Deng, G. Ortiz, A. Poudel, L. Viola, Phys. Rev. B 89,
140507(R) (2014).

\bibitem{J.Li-2009} J. Li, R. L. Chu, J. K. Jain, and S. Q. Shen, Phys. Rev.
Lett. \textbf{102}, 136806 (2009).

\bibitem{W.Li-2011} W. Li, J. Zang, and Y. Jiang, Phys. Rev. B \textbf{84},
033409 (2011).

\bibitem{Y.Y.Zhang-2012} Y. Y. Zhang, R. L. Chu, F. C. Zhang, and S. Q.
Shen, Phys. Rev. B \textbf{85}, 035107 (2012).

\bibitem{Northrup-1982} J. E. Northrup and M. L. Cohen, Phys. Rev. Lett.
\textbf{57}, 154 (1986).

\bibitem{Fan-1989} W. C. Fan and A. Ignatiev, Phys. Rev. B \textbf{40}, 5479
(1989).

\bibitem{Seidel-2009} J. Seidel, et al., Nat. Mater. \textbf{8}, 229 (2009).

\bibitem{Meier-2012} D. Meier, et al., Nat. Mater. \textbf{11}, 284 (2012).

\bibitem{Lee1-1981} D. H. Lee and J. D. Joannopoulos, Phys. Rev. B \textbf{23%
}, 4988 (1981).

\bibitem{YYZhao-1} Y. Y. Zhao, W. Li, and R. B. Tao, Chin. Phys. B \textbf{21%
}, 027302 (2012).

\bibitem{YYZhao-2} Y. Y. Zhao, W. Li, and R. B. Tao, Physica. B \textbf{407}%
, 724 (2012).

\bibitem{Ho-1975} W. Ho, S. L. Cunningham, W. H. Weinberg, and L.
Dobrzynski, Phys. Rev. B 12, 3027 (1975).

\bibitem{Mostoller-1982} M. Mostoller and A. K. Rajagopal, Phys. Rev. B
\textbf{25}, 6168 (1982).
\end{thebibliography}
\end{document}